\documentclass[a4paper,12pt]{article}

\usepackage[english]{babel}
\usepackage[dvips]{graphicx}
\usepackage{amsmath}
\usepackage{amssymb}
\usepackage{overcite}

\begin{document}

\title{Diffusion coefficient for reptation of polymers with kinematic disorder.}
\author{Richard D. Willmann \\
\begin{scriptsize}Institut f\"ur Festk\"orperforschung, Forschungszentrum J\"ulich, 52425 J\"ulich, Germany \end{scriptsize} \\
\begin{scriptsize} email: r.willmann@fz-juelich.de \end{scriptsize} }
\maketitle
\abstract{We give a lower bound on the diffusion coefficient of a polymer chain in an entanglement network with kinematic disorder, which is obtained from an exact calculation in a modified Rubinstein-Duke lattice gas model with periodic boundary conditions. In the limit of infinite chain length we show the diffusive motion of the polymer to be slowed down by kinematic disorder by the same factor as for a single particle in a random barrier model.}
\newpage

\textit{Introduction}: Among the basic problems of polymer science is the derivation of large scale properties of entangled polymers from microscopic properties, such as the molecular weight, which is proportional to the chain length $L$. A scaling argument due to de Gennes \cite{deGennes} predicts for the zero-field diffusion constant  $D(0)$ of a reptating polymer chain that to leading order $D(0) \propto L^{-z}$, $z=2$. Computer simulations \cite{Deutsch,Reiter} and experiments \cite{Lodge1, Lodge2} showed an effective scaling exponent $z$ of 2.28-2.4 for the accessible range of polymer length in contrast to the pure reptation prediction. Reptation based theories accounting for contour length fluctuations (CLF) \cite{Rub} and contraint release (CR) \cite{constraint} (partially) explain this behaviour and predict for increasing chain length a crossover to $z=2$. So far, this region is not experimentally accessible \cite{Lodge2}. In terms of the Rubinstein-Duke (RD) model \cite{Rub,Duk}, which incorporates CLF, it is possible to compute the proportionality constant after the crossover \cite{Koi,Prae} i.e.: $\lim_{L \to \infty}D(0)L^2 = W/(2d+1)$, where $W$ is an elementary hopping rate setting the time scale of defect diffusion and $d$ is the dimensionality of the system environment. Finite size corrections behave as $D(0)L^2-W/(2d+1) \propto L^{- \beta}$, where $1/2 \leq \beta \leq 1$. The experimental relevance of the model is shown in \cite{Exp1,Exp2}. The RD model is an effective model neglecting many effects such as self avoidance of the chain or the short time Rouse dynamics. Moreover, the entanglement network as encountered e.g. in gel electrophoresis is idealized as being regular and static. Real entanglement networks have a random structure whose effects on the motion of the polymer have to be taken into account \cite{Ebert}:
\begin{itemize}
\item Spatial variations of the mobility of the 'defects' of stored length.
\item Locally fluctuating potential energy due to interactions between chain and environment.
\item Entropically favourable regions of low entanglement density.
\item Relaxation of the environment (CR).
\end{itemize}
As many effects are at interplay, it is experimentally impossible to isolate the influence of a single one. However, theoretical considerations and computer simulations can be used to investigate each effect separately. The review \cite{Mut} treats entropic effects and the occurance of 'entropic trapping'. Relaxation of the environment is of minor importance in gels but is considered important in polymer melts \cite{constraint}. \newline
The scope of this communication is the investigation of the influence of kinematic disorder, i.e., disorder reflecting varying defect mobility without affecting the equilibrium configuration of the chain. In \cite{Ebert} an analysis of Monte Carlo data for a polymer with kinematic disorder, i.e., spatially varying mobility of defects, is performed, which shows reptation dynamics to prevail. Being based on computer simulations and thus short chains, only speculations about the limit of infinite chain length are possible. It is argued that in this limit the diffusion constant still scales as 
\begin{equation}
D(0)L^{2}=c,
 \label{Gleichung} 
\end{equation}
where the constant of proportionality $c$ might be some average of the hopping rates $W$. Using a  modified RD model we can give partial confirmation to this conjecture by rigorously proving a lower bound $D^{per}$ on $D(0)$, which yields (\ref{Gleichung}). Moreover, we explicitly calculate the constant $c$ and thus show, how the disorder changes the coefficient. In the limit of infinite chain length $D^{per}L^{2}=1/<1/W>$.
 \newline \textit{Definition of the model}: In the RD model, the entanglement network is represented as a cubic lattice, the lattice constant being equal to the mean entanglement length. A string of $L+1$ 'reptons', i.e., sections with a length of the lattice constant, represents the polymer. The repton dynamics is as follows:
\begin{itemize}
\item [a)] Each cell occupied by the chain must contain at least one repton to ensure connectivity of the chain.
\item [b)] End reptons can move to adjacent cells provided rule a) is not violated.
\item [c)] Interior reptons can move to cells occupied by the neighbouring reptons if allowed by a). This ensures the dynamics to be reptation.
\end{itemize}

\begin{figure}[h]
\begin{center}
\includegraphics[totalheight=8cm,keepaspectratio]{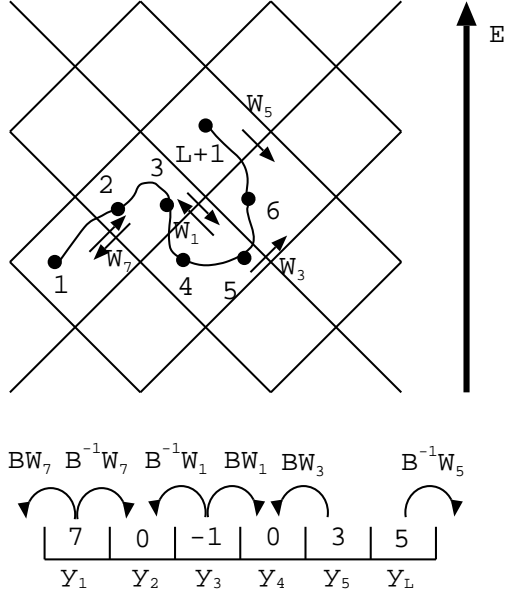}
\caption{Two dimensional representation of  a network with a polymer chain and mapping to the lattice gas model. Arrows show possible moves.}
\end{center}
\end{figure}

Considering kinematic disorder each boundary between cells has assigned to it an individual hopping rate for a repton crossing in any direction . We assume there to be $\sigma \in \mathbb{N}$ possible rates $W_{\alpha}$, each occuring with probability $f(W_{\alpha})$ throughout the network (Fig. 1). We demand that for the distribution $f(W_{\alpha})$ the disorder averages $\langle 1/W \rangle$ and $\langle 1/W^2 \rangle$ are finite. \newline
The RD model is a model for electrophoresis. The electric field $E$ points along a body diagonal of the cubes in the lattice and each repton carries a charge. By local detailed balance, this modifies the rates for reptons crossing cell boundaries by a factor $B^{\pm 1}$ depending on it moving along (+) or against (-) the field, where $B=\exp(E/2)$ \cite{Duke2}. The configuration of the chain can by rephrased as a one dimensional lattice gas model with $L$ sites by considering the links between reptons with respect to $E$. Links between reptons in the same cell are represented as '0' (vacancy), those which are oriented along (against) the field and across a cell boundary with rate $W_{\alpha}$ assigned to it as particles of type '$\alpha$' ('$-\alpha$'). Thus, the chain conformation is represented by $L$ 'pseudospins' $y_1$ to $y_L$ (Fig. 1). Rule c) for the repton dynamics enforces the lattice gas dynamics to be as for an exclusion process: In the bulk particles of sort '$\pm \alpha$' hop to the left with rate $B^{\pm 1} W_{\alpha}$ and to the right with rate $B^{\mp 1} W_{\alpha}$, where each site can be occupied by at most one particle. The end dynamics in the lattice gas picture needs some care: Assuming $y_1$ ($y_L$) to be non zero, the only possible move is the retraction of the end repton to the cell occupied by it's neighbour (rule a)). This retraction, being an annihilation event in the lattice gas picture, happens with the same rate as the respective move in the bulk. Assuming $y_1$ ($y_L$) to be zero, the end repton can, according to rule b), move to any of the $2d$ adjacent cells. For half of these the move leads to links being along the field direction, the other half against it. The probability of the chosen move leading to the crossing of a cell boundary with rate $W_{\alpha}$ being assigned to it is $f(W_{\alpha})$. Thus the move of the repton, being a creation event in the lattice gas picture leads to $y_1$ ($y_L$) changing from '0' to '$\pm \alpha$' with rate $B^{\mp1}f(W_{\alpha})W_{\alpha}d$ ($B^{\pm 1}f(W_{\alpha})W_{\alpha}d$). This choice of boundary dynamics is on average correct, but neglects the actual local structure of the network \cite{paper}. Yet to define is the motion of the centre of mass position $x$ in terms of the lattice gas model:
\begin{itemize}
\item Particle type '$\alpha$' moving to the right (left) decreases (increases) $x$ by $1/(L+1)$, as this is equivalent to a repton moving downward (upward). As there are $L+1$ reptons each contributes $1/(L+1)$ to the centre of mass position.
\item Particle type '$-\alpha$' moving to the right (left) increases (decreases) $x$ by $1/(L+1)$.
\end{itemize}
In the RD-model, discriminating links between reptons along ('$\alpha$') and against ('$- \alpha$') the field direction, which is an arbitrarily chosen direction in space, allows for following the transport of stored length along this direction. Thus the zero field diffusion constant along this space direction can be calculated \cite{Koi,Prae}, which immediately yields the 3-dimensional diffusion constant as diffusion at zero field is isotropic. This is in contrast to the original Rubinstein model \cite{Rub}, which allows only for the calculation of the curvilinear diffusion constant along the contour of the tube within the model and requires additional assumptions to relate it to the 3-dimensional diffusion constant.\\
\textit{Relation of open and periodic system}: Calculations proceed along analogous lines as in \cite{Koi,Prae}. The adaption to the disordered system is straightforward, details will be presented in a forthcoming publication \cite{paper}. Using detailed balance, we calculated the stationary state $P^*_{open}(0)$ at zero field with respect to $H_{open}$. It is a product measure and the probability of finding a configuration $\mathbf{y}=(y_1,..,y_L)$ is given by
\begin{equation}
P^*_{open}(0)=\prod_{i=1}^L \tilde{P}(y_i) \text{ with } \tilde{P}(y_i) = \left\{ \begin{array}{cl} 1/(2d+1) & \text{for } y_i=0. \\ f(W_{\alpha})d/(2d+1) & \text{for } y_i=\pm \alpha. \end{array} \right.
  \label{stat} 
\end{equation}
The shape of the chain only depends on the signs of the $y_i$. According to (\ref{stat}) the probability for $y_i=0$ is $1/(2d+1)$, for $y_i$ being positive (negative) $d/(2d+1)$. These probabilities are as for the original RD model, implying that our kind of disorder leaves the equilibrium conformation of the chain unaffected, as in \cite{Ebert}. \newline
For the same bulk dynamics, but periodic boundary conditions, it turns out that $P^*(0)$ is at zero field also a stationary state. This enables comparing the influence of the boundary terms of the stochastic generators on the diffusion constant as in \cite{Diplom} and we can prove that $D_L(0) \geq D_{L+1}^{per}$, where $D_L(0)$ means $D(0)$ for a lattice gas with $L$ sites. $D_{L+1}^{per}$ is the centre of mass diffusion constant for a lattice gas of $L+1$ sites and periodic boundary conditions, where the centre of mass variable $x$ depends on particle moves in the lattice gas as was stated above for the open system. In the following we calculate $D_{L+1}^{per}$ to leading order and thus provide a lower bound on $D(0)$. As it is undisputed that disorder slows down diffusion compared to the naive approximation $c=<W>$, this is the physically relevant bound. \newline 

\begin{figure}[h]
\begin{center}
\includegraphics[scale=1.25]{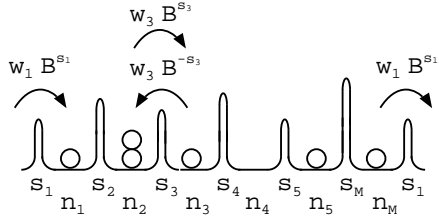}
\caption{Zero range particles moving in a random barrier energy landscape.}
\end{center}
\end{figure}

\textit{Zero range picture}: Dealing with a periodic system it is convenient to use the following alternative point of view: Instead of characterizing the system by $\mathbf{y}=(y_1,..,y_L)$, it is possible to use the sequence $\mathbf{s}=(s_1,..,s_M)$ of the signs of those $y_i$, which are non zero and have rates $\mathbf{w}=(w_1,..,w_M)$, $w_j \in \{W_1,..,W_{\sigma}\}$, and $\mathbf{n}=(n_1,..,n_M)$, where $n_j$ equals the number of $y_i=0$ between $s_j$ and $s_{j+1}$. Here, $s_{M+1} \equiv s_1$. $n_j$ can be understood as counting particles of zero interaction range at site $j$. To make this alternative picture consistent with the lattice gas picture, we require $\sum_{j=1}^M n_j = L-M$. Also the dynamics of the $n_j$ is thus fixed: The configuration $(..,n_j,n_{j+1},..)$ changes to $(..,n_j-1,n_{j+1}+1,..)$ with rate $B^{s_{j+1}}w_{j+1}$ and to $(..,n_j+1,n_{j+1}-1,..)$ with rate $B^{-s_{j+1}}w_{j+1}$. This means that the random hopping rates as well as the $s_j$ are not assigned to individual particles, but to bonds between sites in the zero range (zr) picture \cite{BFL}. At $E=0$ the zr-particles move as in a random barrier energy landscape (Fig. 2). A zr-particle hopping to the right across a bond with $s_j>0$ ($s_j<0$) increases (decreases) the centre of mass position by $1/L$. Conversely, a zr-particle hopping to the left across a bond with $s_j>0$ ($s_j<0$) decreases (increases) the centre of mass position by $1/L$. 
Regarding a periodic system, the phase space is non ergodic, as in the zr picture neither the number, nor the order of the $s_j$ and $w_j$ can be changed. Therefore, the phase space constists of connected subspaces, called 'channels' \cite{Koi}, being characterized by $\mathbf{s}$ and $\mathbf{w}$. For obtaining the expectation value of the centre of mass drift velocity for the full phase space $\bar{v}$, at first the expectation value of the centre of mass drift velocity for each channel $<v>$ has to be calculated. Subsequently averaging over the expectation values for each channel yields $\bar{v}$. To compute $D^{per}_L$ we will employ the Einstein relation $D^{per}_L=1/L(d\bar{v}/dE)_{E=0}$. \newline
\textit{Calculations for individual channels}: Expressing the dynamics for the zr-particles as stated above by a stochastic generator $H_{zr}^{s,w}$, we computed the stationary state $P_{zr}^{s,w}$ for arbitrary $E$. The use of a product measure ansatz leads to a recursion relation yielding the following steady state probability for a configuration $\mathbf{n}=(n_1,..,n_M)$ \cite{paper}:
\begin{equation} 
P_{zr}^{\mathbf{s},\mathbf{w}}(\mathbf{n})= \prod_{j=1}^M z_j^{n_j} \big / \sum_{\mathbf{n}}' \prod_{j=1}^M z_j^{n_j}.
\label{pzr}
\end{equation}
The primed sum means summing with the constraint $\sum_{j=1}^M n_j=L-M$ and 
\begin{equation} z_j=\sum_{i=1}^M 1/(\exp(s_{j+i}E/2)w_{j+i})\prod_{k=1}^{i-1}\exp(-s_{j+k}E).
\end{equation}
The drift velocity is in the lattice gas picture given by the difference of currents of particles with $y_i<0$ and $y_i>0$: $<v>=<j^--j^+>$. This translates into the zr picture as the current of zr-particles across bonds with $s_j>0$ minus the one across bonds with $s_j<0$. Due to using the Einstein relation, only first order terms in an expansion of $<v(\mathbf{s},\mathbf{w})>$ into $E$ contribute to $D^{per}_L$. The quantity $z_i$ as occuring in (\ref{pzr}) can then be calculated as $z_i=\sum_{j=1}^{M} 1/w_{i+j}=z$. This facilitates evaluating $<v(\mathbf{s},\mathbf{w})>$, which using the quantum Hamiltonian formalism \cite{Schuetz} and a treatment as in \cite{Koi} yields
\begin{equation}
<v(\mathbf{s},\mathbf{w})>=E S^2  \frac{L-M}{L(L-1)}  \frac{1}{z} + o(E^2)
\end{equation}
with $S=\sum_{j=1}^M s_j$. \newline
\textit{Averaging over the channels}: To obtain $\bar{v}$, an average of $<v>$ over the channels has to be performed, where each channel has to be weighted such that in zr and lattice gas picture corresponding configurations have equal weight in the stationary state. $\Psi(\mathbf{s},\mathbf{w})$ is the weight factor of the channels as in \cite{Koi,Diplom}, which is modified by the disorder to
\begin{equation}
\Psi(\mathbf{s},\mathbf{w})=\frac{d^M}{(2d+1)^L}{L \choose M}  \prod_{j=1}^M f(w_j).
\end{equation}
Thus $\bar{v}$ is to first order in $E$ given by
\begin{equation}
\begin{split}
\bar{v} & =  \sum_M \sum_{\mathbf{s}=(s_1,...,s_M)} \sum_{\mathbf{w}=(w_1,...,w_M)} \Psi(\mathbf{s},\mathbf{w})  <v(\mathbf{s},\mathbf{w})> \\
& =  \overline{v_o}  < \frac{1}{z}>,
\end{split}
\end{equation}
where $<1/z>$ is the average of $z$ with respect to the distribution $f(W_{\alpha})$ and 
$\overline{v_o}$ is the result for the RD model without disorder. For $M \rightarrow \infty$ the restrictions on $f(W_{\alpha})$ allow invoking the central limit theorem, which yields: $<1/z>=1/(M<1/W>)$, leading, when employing the Einstein relation, to the following result in the limit of infinite chain length:
\begin{equation}
D_L^{per}L^2=\frac{1}{(2d+1)} \frac{1}{<1/W>}.
\end{equation}
Comparing to the result in \cite{Koi} for the ordered case RD model, $D_L^{per}L^2=1/(2d+1)$ reveals the remarkable result that in this limit the center of mass diffusion in the RD model on a ring is slowed down by kinematic disorder in the same manner as the single particle diffusion constant in a random barrier model.
We remark that for the ordered RD model, $D_{L+1}^{per}$ and $D_L(0)$ are equal to leading order as shown in \cite{Prae} by a variational statement for $D(0)$. This variational technique is also applicable to the RD model modified by kinematic disorder \cite{paper} and shows, that the naive approximation $c=<W>$ provides an upper bound for $D(0)$. We performed Monte Carlo simulations of the model with open boundary conditions with various distributions $f(W)$ and chain length up to $L=40$. Apart from quickly decaying finite size effects, which depend on $f(W)$, the results clearly indicate that also for the RD model with kinematic disorder open and periodic system have to leading order the same zero field diffusion constant and therefore indeed $c=1/<1/W>$ yields the correct asymptotic behaviour.
\newline
Gunter M. Sch\"utz is gratefully acknowledged for posing the problem and many helpful discussions.

\end{document}